# Direct Evidence for Multiferroic Magnetoelectric Coupling in $0.9BiFeO_3$-$0.1BaTiO_3$

Anar Singh[1], Vibhav Pandey[2], R. K. Kotnala[2] and Dhananjai Pandey[1*]

[1]School of Materials Science and Technology, Institute of Technology, Banaras Hindu University, Varanasi -221305, India
[2]National Physical Laboratory, K. S. Krishnan Marg, Pusa, New Delhi-110012, India

Magnetic, dielectric and calorimetric studies on $0.9BiFeO_3$-$0.1BaTiO_3$ indicate strong magnetoelectric coupling. XRD studies reveal a very remarkable change in the rhombohedral distortion angle and a significant shift in the atomic positions at the magnetic Tc due to an isostructural phase transition. The calculated polarization using Rietveld refined atomic positions scales linearly with magnetization. Our results provide the first unambiguous evidence for magnetoelectric coupling of intrinsic multiferroic origin in a $BiFeO_3$ based system.

[*]Author for correspondence; electronic mail: dpandey_bhu@yahoo.co.in

The current surge of interest in multiferroic materials showing magnetoelectric coupling due to the presence of both magnetic and ferroelectric ordering is fuelled by both the potential technological applications and the underlying new physics [1-3]. The magetoelectric coupling (ME) provides an additional degree of freedom in the designing of actuators, sensors, and data storage devices [3]. From the fundamental physics point of view, the coexistence of ferroelectric and magnetic order is contra-indicated [1], as the former requires empty 'd' orbitals for the off-centre displacement of cations responsible for ferroelectricity [4] while the latter results from partially filled d orbitals . The coexistence of such mutually exclusive phenomena in some compounds has been attributed to new mechanisms of ferroelectricity, such as lone pair stereochemical activity of the 'A' site cation [1-3,], loss of inversion symmetry due to spiral magnetic ordering [1-3], and change in geometrical arrangement of ions [1,2]

$BiFeO_3$ (BF) is one of the most extensively investigated multiferroic magnetoelectric compounds in which the Bi 6s lone pair electrons are believed to be responsible for ferroelectricity [1, 2 and references there in] while partially filled 'd' orbitals of Fe lead to magnetic ordering. $BiFeO_3$ is unique amongst various multiferroic magnetoelectrics, as its ferroelectric and magnetic transition temperatures are well above the room temperature raising the possibility of potential devices based on magnetoelectric coupling operating at the room temperature. Bulk $BiFeO_3$ has got a rhombohedrally distorted perovskite structure in the R3c space group [5] with a ferroelectric Tc of ~ 1103 K [6]. It shows G-type antiferromagnetic spin configuration along $[001]_h$ direction with a long range (periodicity ~ 620 Å) cycloidal spin structure incommensurate with the lattice along $[110]_h$ direction of the hexagonal unit cell of the rhombohedral structure below ~ 643 K [7, 8]. This long period modulated magnetic structure leads to cancellation of net macroscopic magnetization and hence inhibits the observation of linear magnetoelectric effect [9]. However, spin flop has been reported under electrical poling [10]. The incommensurate cycloidal spin structure can be suppressed with the application of high magnetic fields >18 T [11], and probably by chemical substitutions too [1-3]. A significant magnetization (1 $\mu_B$/unit cell) and strong magnetoelectric coupling has been observed in epitaxial thin films, suggesting that the spiral spin structure could possibly be suppressed in thin films also [12].

The observation of an anomaly in the dielectric constant at the magnetic transition temperature is taken as an evidence of multiferroic magnetoelectric coupling [1-3]. However, in sintered polycrystalline samples such a dielectric anomaly at the magnetic transition temperature may not necessarily result from multiferroicity of the bulk sample only, as it may have significant contributions from space charge effects at the interfacial layers (i.e. grain boundaries, grain-electrode interfaces) of different resistivities [13]. For an unambiguous evidence of multiferroic magnetoelectric coupling at the magnetic transition temperature in ceramic samples, it is therefore essential to separate out various space charge contributions from the bulk contribution. Here, we present unambiguous evidence for magnetoelectric coupling of intrinsic multiferroic origin by temperature dependent dielectric, magnetic, calorimetric and X-ray diffraction (XRD) studies on a $BaTiO_3$ substituted $BiFeO_3$ composition, $0.9BiFeO_3$-$0.1BaTiO_3$ (BF-0.1BT). We use complex impedance spectroscopic technique (Cole-Cole plot) to identify the frequency domain above which the dominant contribution to the dielectric anomaly at the magnetic transition temperature is due to the bulk (grain) phase only. We confirm this bulk magnetoelectric coupling of intrinsic multiferroic origin by temperature dependent XRD studies which not only reveal pronounced anomaly in the rhombohedral distortion angle at the magnetic transition temperature due to strong magnetoelastic coupling, but also show significant shift in the atomic positions below Tc characteristic of an isostructural phase transtion. The calculated polarization (P) obtained from the structural data assuming ionic model is shown to scale with magnetization (M) thereby providing first direct evidence for magnetoelectric coupling of multiferroic origin via magnetoelastic effect in $BiFeO_3$ based systems.

The synthesis of monophasic samples of $BiFeO_3$ and its solid solutions in bulk continues to be a challenging task. One of the commonly reported impurity phases is $Bi_2Fe_4O_9$ [$(Bi_2O_3)2.(Fe_2O_3)$] [7], which is an Fe-rich phase and can be removed by leaching [14]. However, notwithstanding the leaching out of the impurity phase after its formation, the very formation of this impurity phase would obviously disturb the overall stoichiometry of the final perovskite phase and hence the overall physical properties. We have minimised this impurity phase content in our BF-0.1BT samples to a nearly negligible level (<1%) by controlling the loss of $Bi_2O_3$ during sintering due to its significantly high vapour pressure at the sintering temperature. Stoichiometric mixture of analytical reagent grade $Bi_2O_3$ (99.5%), $Fe_2O_3$ (99%), $BaCO_3$ (99%), and $TiO_2$ (99%) was calcined at 1033 K for 8 hours in open alumina crucible and sintered at 1173 K for 1 hour in closed alumina crucible with calcined powder of the same composition used as spacer powder.

Powder X-ray diffraction (XRD) studies were carried out using an 18KW Cu-rotating anode based Rigaku (Tokyo, Japan) powder diffractometer operating in Bragg-Brentano geometry and fitted with a curved crystal graphite monochromator in the diffraction beam and a high temperature attachment. The temperature dependent dielectric and impedance measurements were carried out in the temperature range 300 to 750 K at a heating rate of 1 K/min using a Nova Control (Alpha-A) high performance frequency analyzer. The M-H loop at room temperature and temperature dependent magnetization measurements in the temperature range 300 to 725 K at a low magnetic field of 500 Oe were carried out using a vibrating sample magnetometer (VSM-7305, Lakeshore) at a heating rate of 5 K/min. For differential scanning calorimetry (Mettler DSC827$^e$), a heating rate of 1 K/min was used.

The inset of Fig.1(a) shows the magnetization (M)-magnetic field (H) loop of BF-0.1BT at room temperature which indicates weakly ferromagnetic nature, unlike pure $BiFeO_3$ which does not show [15] any M-H loop. This hysteresis loop disappears above 650 K. It cannot therefore be attributed to the presence of impurity phases like $Bi_2Fe_4O_9$ or $\gamma$-$Fe_2O_3$ whose magnetic transition temperatures are 260 K [16] and 850 K [17], respectively. Our results also suggest that the value of Tc = 255 K reported earlier in BF-0.1BT [18] is not due to the fereromagnetic phase of BF-0.1BT but is most likely due to $Bi_2Fe_4O_9$. The observation of M-H loop with a small but non-zero spontaneous magnetization in BF-0.1BT samples suggests that the cycloidal spin structure of pure $BiFeO_3$ ceramic may be getting suppressed in the BF-0.1BT solid solution. Fig.1 (a) shows the temperature dependence of the inverse of dc magnetization at H = 500 Oe in the temperature range 300 to 725 K. The extrapolation of the straight line fit to the high temperature region gives a magnetic transition temperature Tc ~ 648 K. This transition temperature is in agreement with the Tc obtained by DSC studies which reveal an anomaly at 645 K, as can be seen from Fig.1 (a). We also observe a pronounced anomaly in the dielectric constant (ε′) near the magnetic transition temperature, as can be seen from Fig.1 (b). However, the temperature ($T_m^{/}$) corresponding to the peak in ε′(T) shows considerable frequency dispersion. It shifts from 532 K at 1 kHz to 637 K at 300 kHz on increasing the measuring frequency, with a concomitant decrease in the peak value of the dielectric constant at $T_m^{/}$, mimicking a relaxor ferroelectric behavour. The relaxation time (τ) corresponding to the frequency dependent $T_m^{/}$ follows Arrhenius dependence with two different slopes and not Vogle-Fulcher law observed in canonical relaxor ferroelectric systems like $Pb(Mg_{1/3}Nb_{2/3})O_3$ [19]. This can be seen from the inset of Fig.1(b) which depicts the lnτ vs 1/T plot. The first slope corresponds to frequencies ≤300 kHz whereas the second one for frequencies higher than 300 kHz with activation energies of 1.5 and 2.48 eV, respectively. It is interesting to note from Fig. 1(b) that for frequencies >300 kHz, the $T_m^{/}$ shows less frequency dispersion and it nearly coincides with the magnetic transition temperature Tc~648 K. In order to verify the presence of intrinsic (bulk) as well as the space charge contributions to our measured values of the dielectric constant, we carried out impedance spectroscopic analysis.The Cole-Cole plot for the complex impedance (Z* = Z′ -iZ″) of BF-0.1BT in the 100 Hz to 1 MHz frequency range at 575 K clearly reveals the presence of three overlapping semicircular arcs (see Fig.2) which are attributable to contributions from the grain, grain boundary and electrode-grain interfaces in the decreasing order of measuring frequencies [20].

What could be the origin of the widely different resistivities of grains and grain boundaries leading to the characteristic semicircular arcs in Fig.2? It is well known that the equilibrium concentration of oxygen vacancies in $ABO_3$ type perovskites sintered in reducing or low oxygen pressure environments can be sufficiently large [21]. Since our BF-0.1BT samples were sintered in closed crucibles to prevent $Bi^{3+}$ losses, the oxygen partial pressure inside the crucibles may be low, as a result of which there would be loss of oxygen during sintering in BF-0.1BT. This loss can be written in the Kröger-Vink notation as follows:

$$O_0 \Leftrightarrow 1/2\, O_2 + V_0^{\cdot\cdot} + 2e^{/} \quad (1)$$

Electrons released due to oxygen vacancy $V_0$ may be captured by $Fe^{3+}$ of BF-0.1BT, leading to its reduction to $Fe^{2+}$. The presence of $Fe^{2+}$ and $Fe^{3+}$ ions would lead to hopping of electrons which increases conductivity. During the cooling down period after sintering, the reverse process of reaction (1) occurs since the equilibrium concentration of vacancies decreases. But due to the insufficient time available during cooling down and also the falling furnace temperature, the reoxidation process gests restricted to the grain boundaries only [22]. Thus, the grain boundaries regain their insulating properties but the grains remain semiconducting due to the presence of oxygen vacancies. The hopping charge carriers in the grains would be stopped by the insulating grain boundaries, leading to enormously large space charge contributions to the measured dielectric constant. A similar effect occurs at the conducting electrode-semiconducting grain interfaces also which would further enhance the measured dielectric constant. The two space-charge contributions are over and above the intrinsic dielectric constant of the grains (bulk). Since the interfacial polarization processes have low relaxation times, their contributions decrease drastically above 300 kHz in our BF-0.1BT samples. Thus, the significantly reduced frequency dispersion of $T_m^{/}$ for frequencies greater than 300 kHz in Fig.1 (b) is due to the predominantly bulk (grain) contributions linked with magnetoelectric coupling of intrinsic multiferroic origin. It is, however, interesting to note that even the interfacial contributions to the total dielectric constant are more pronounced near the magnetic transition temperature suggesting the presence of magnetoelectric coupling of non-multiferroic origin as well.

The magnetoelectric coupling of intrinsic multiferroic origin discussed in the foregoing was further confirmed through the temperature dependent structural studies. Powder XRD studies as a function of temperature reveal that the structure of BF-0.1BT remains rhombohedral in the R3c space group across the magnetic Tc atleast upto ~ 875 K. However, the rhombohedral angle, as obtained by Rietveld refinement at each temperature in the range 300 to 875 K, shows clear anomaly at Tc (see Fig. 3) similar to that reported in pure $BiFeO_3$ [23]. Further, the positional coordinates of all the atoms in the asymmetric unit cell of R3c space group shift significantly below Tc (see Fig. 4) and, as shown elsewhere [24], the shifts are consistent with one of the irreps of the R3c space group. The Δz shift in Bi position at 300 K (i.e. below magnetic Tc) and 700K (i.e. above magnetic Tc) is ~0.038 Å which is comparable to shift in $BaTiO_3$ [25]. Since Bi is largely responsible for the ferroelectricity of $BiFeO_3$, the large change in its position in the magnetic phase confirms coupling of ferroelectric (P) and magnetic (M) order parameters. As the ferroelectric space group does not change across Tc~648 K, these observations suggest an isostructural phase transition similar to that reported in $YMnO_3$ and $LuMnO_3$ [26] due to magnetoelastic effect. We have calculated the polarization (P) using the positional coordinates and the nominal valence of the ions involved. The spontaneous polarization obtained by us (~57 $\mu C/cm^2$ at 300 K) along $[001]_h$ (or $[111]_R$) using this simplified model is not very much different from the experimental value (~60 $\mu C/cm^2$) reported on good quality insulating $BiFeO_3$ crystals [15]. The polarization so obtained not only increases in the ferromagnetic state but also scales with magnetization (M) (see Fig. 5) providing direct evidence for the intrinsic magnetoelectric coupling of multiferroic origin. The observation of linear magnetoelectric coupling (see Fig. 5) in BF-0.1BT also suggests that the magnetic spiral structure of $BiFeO_3$ is suppressed as a result of $BaTiO_3$ substitution, as only quadratic magnetoelectric coupling is expected for the spiral magnet [3]. To conclude, our results provide first unambiguous and direct evidence for magnetoelectric coupling of intrinsic multiferroic origin in a $BiFeO_3$ based system.

One of the authors (Anar Singh) acknowledges the award of Junior Research Fellowship of UGC, India.

**Figure Captions:**

**Figure 1.** Temperature dependence of (a) the inverse of the magnetic susceptibility (dotted line) measured at 500 Oe and background subtracted heat flow (continuous line), (b) the real part of the dielectric constant at various selected frequencies (1 to 8 correspond to 1, 10, 50, 100, 300, 500, 700, 1000 kHz). The insets to Figures (a) and (b) depict the M-H hysteresis loop at 300 K and the Arrhenius plot for the polarization relaxation time, respectively.

**Figure 2.** Cole-Cole plot of complex-impedance ($Z^{*}= Z'-iZ''$) at 575 K in the frequency range 100 Hz to 1 MHz. Symbols are experimental data points, while the semicircular arcs are the fitted depressed circles.

**Figure 3.** Temperature dependence of the variation of the rhombohedral distortion angle and the unit cell volume.

**Figure 4.** Temperature variation of the positions of the atoms in the asymmetric unit cell.

**Figure 5.** Variation of the polarization with temperature (filled circles) and polarization with magnetization (open circles).

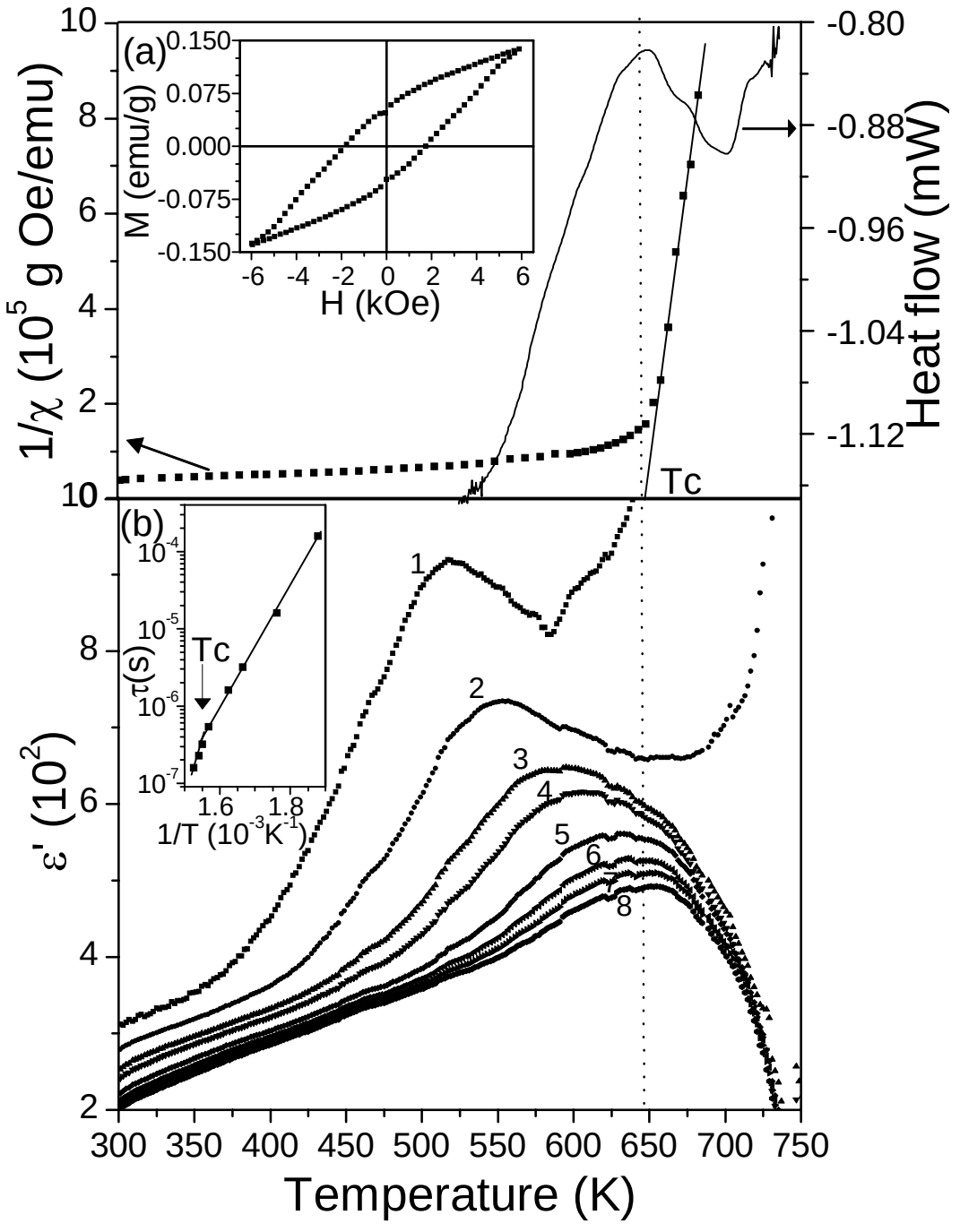

**Fig. 1**

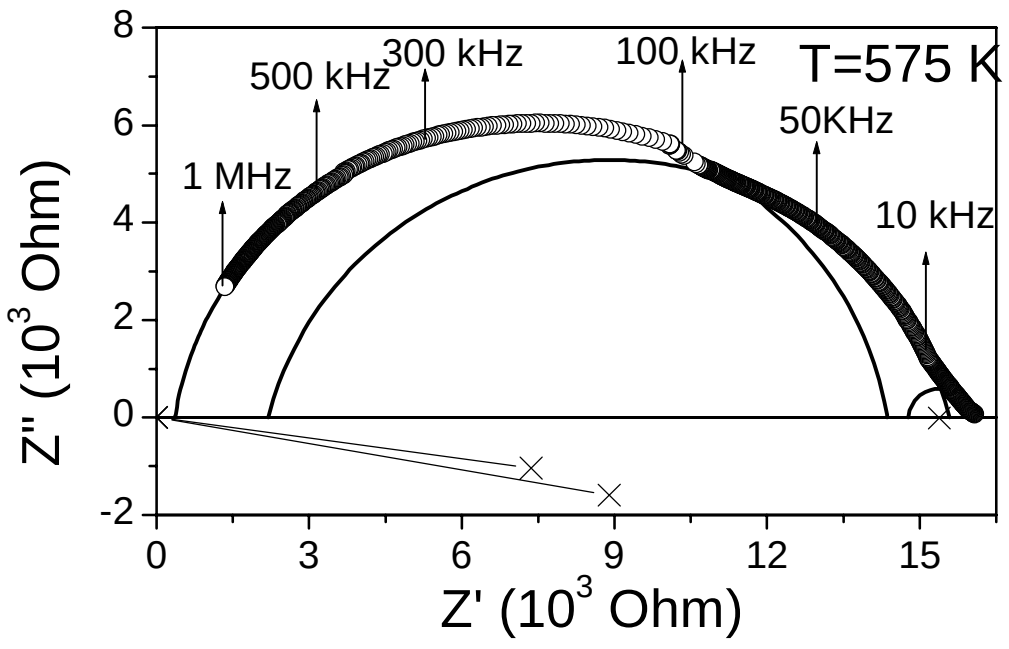

**Fig. 2**

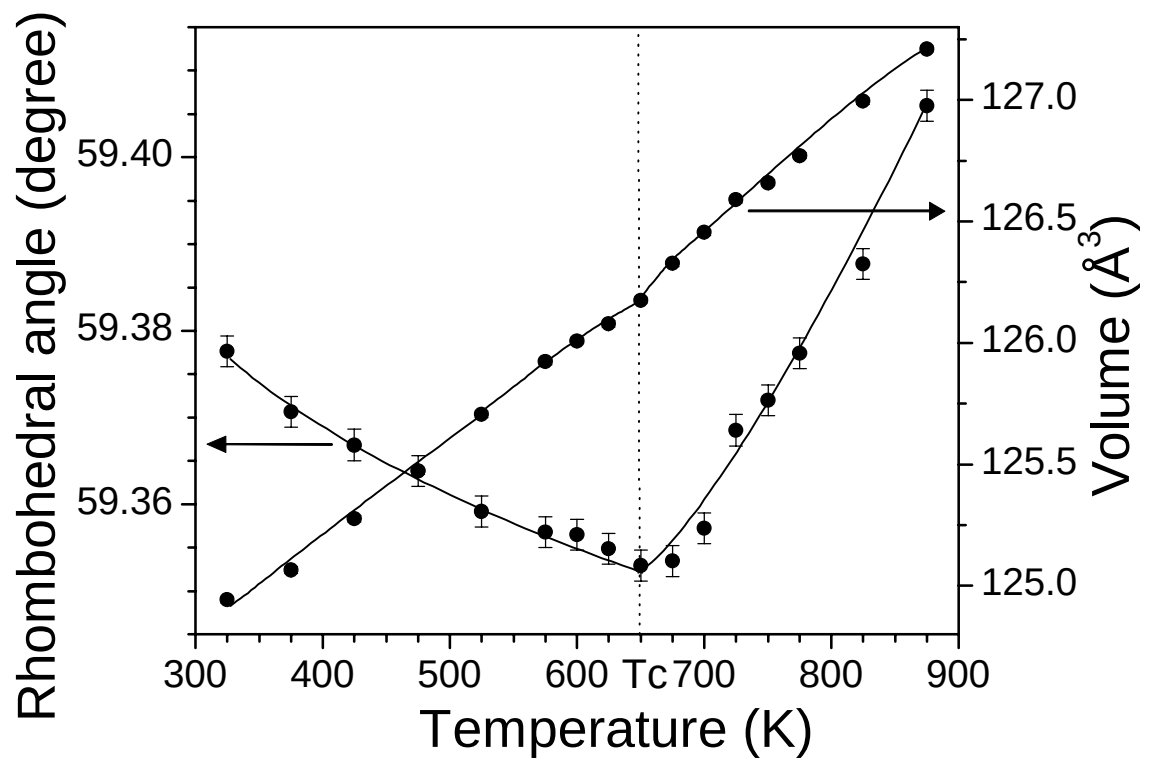

**Fig. 3**

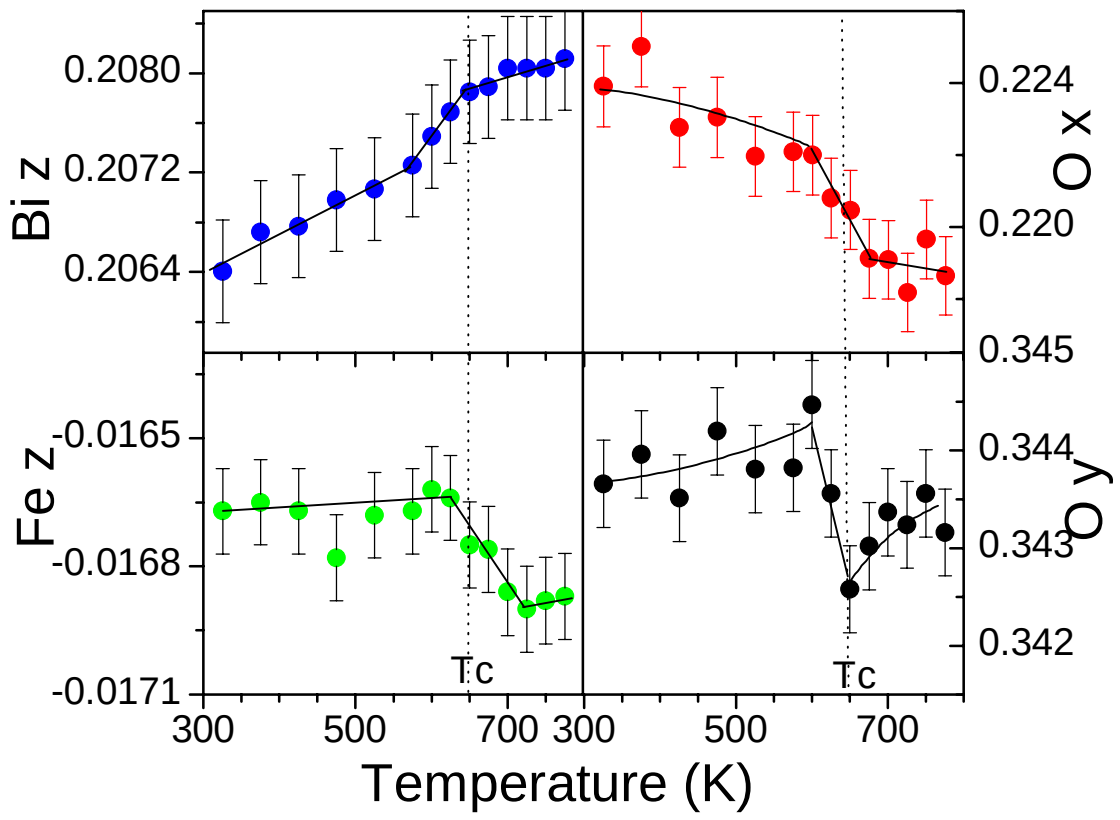

**Fig. 4**

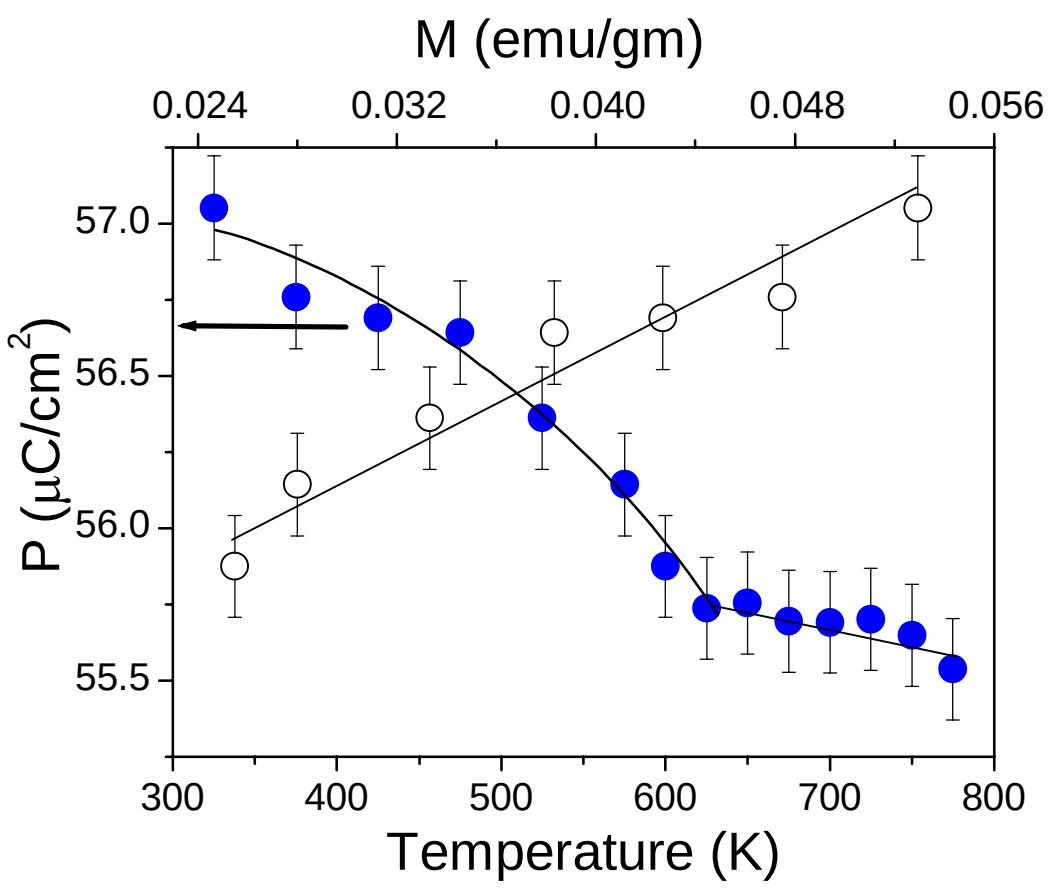

**Fig. 5**